\documentclass[twocolumn]{aastex62}

\usepackage{amsmath}
\usepackage{amssymb}
\usepackage{empheq}
\usepackage{mathrsfs}
				
\usepackage{amssymb}

\newcommand{\G}{\mathcal{G}}
\newcommand{\M}{M_{\star}}
\newcommand{\Msun}{M_{\odot}}
\newcommand{\R}{\mathcal{R}}
\newcommand{\Ham}{\mathcal{H}}
\newcommand{\mn}{m_{\rm{N}}}
\newcommand{\an}{a_{\rm{N}}}
\newcommand{\en}{e_{\rm{N}}}
\newcommand{\Mn}{M_{\rm{N}}}
\newcommand{\order}{\mathcal{O}}

\begin{document}


\title{The Stability Boundary of the Distant Scattered Disk}

\author{Konstantin Batygin}
\affiliation{Division of Geological and Planetary Sciences California Institute of Technology, Pasadena, CA 91125, USA}

\author{Rosemary A. Mardling}
\affiliation{School of Physics and Astronomy, Monash University, Victoria, 3800, Australia}

\author{David Nesvorn{\'y}}
\affiliation{Department of Space Studies, Southwest Research Institute, 1050 Walnut St., Suite 300, Boulder, CO 80302, USA}

\begin{abstract}
The distant scattered disk is a vast population of trans-Neptunian minor bodies that orbit the sun on highly elongated, long-period orbits. The orbital stability of scattered disk objects is primarily controlled by a single parameter -- their perihelion distance. While the existence of a perihelion boundary that separates chaotic and regular motion of long-period orbits is well established through numerical experiments, its theoretical basis as well as its semi-major axis dependence remain poorly understood. In this work, we outline an analytical model for the dynamics of distant trans-Neptunian objects and show that the orbital architecture of the scattered disk is shaped by an infinite chain of exterior $2:j$ resonances with Neptune. The widths of these resonances increase as the perihelion distance approaches Neptune's semi-major axis, and their overlap drives chaotic motion. Within the context of this theoretical picture, we derive an analytic criterion for instability of long-period orbits, and demonstrate that rapid dynamical chaos ensues when the perihelion drops below a critical value, given by $q_{\rm{crit}}=\an\,\big(\ln((24^2/5)\,(\mn/\Msun)\,(a/\an)^{5/2})\big)^{1/2}$. This expression constitutes an analytic boundary between the ``detached" and actively ``scattering" sub-populations of distant trans-Neptunian minor bodies. Additionally, we find that within the stochastic layer, the Lyapunov time of scattered disk objects approaches the orbital period, and show that the semi-major axis diffusion coefficient is approximated by $\mathcal{D}_a\sim(8/(5\,\pi))\,(\mn/\Msun)\,\sqrt{\G\,\Msun\,\an}\,\exp\big[-(q/\an)^2/2\big]$. We confirm our results with direct $N-$body simulations and highlight the connections between scattered disk dynamics and the Chirikov Standard Map. Implications of our results for the long-term evolution of minor bodies in the distant solar system are discussed.
\end{abstract}

                              

\section{Introduction} \label{sec:intro}

Among the various sub-populations of the icy debris that comprise the Kuiper belt, the most prominent -- both in terms of mass and radial extent -- is the scattered disk. A remnant of Neptune's early outward migration \citep{Nesvorny2018REV}, the scattered disk is largely made up of eccentric, low-inclination orbits that ``hug" the orbit of Neptune, maintaining a perihelion distance slightly above $q\gtrsim30\,$AU \citep{2020tnss.book...25M}. Interesting in its own right, the orbital architecture of the distant scattered disk is especially distinctive, as it provides an observational handle on the gravitational processes that have sculpted the outermost reaches of the solar system \citep{2004ApJ...617..645B, 2010ARA&A..48...47A, 2019PhR...805....1B, 2021arXiv210501065C}.

The dynamics of scattered disk objects (SDOs) have been studied in considerable detail over the past two and a half decades, and the general characteristics of their long-term evolution are relatively well understood (see \citealt{Saillenfest2020} and the references therein). Crudely speaking, objects with perihelion distance small enough to strongly interact with Neptune experience chaotic diffusion and eventually become Centaurs\footnote{Centaurs are broadly defined as objects with perihelion distance or semi-major axis that fall between the orbits of Jupiter and Neptune.}, or leave the solar system altogether. To be more precise, the survival probability of a chaotic SDO over the age of the sun is about $1\%$ \citep{Gomes2008}. Conversely, objects with large perihelia -- often referred to as the ``detached" population -- are immune to strong Neptune-induced perturbations, and simply orbit the sun on slowly precessing Keplerian orbits.

Today, orbital integration of scattered disk objects does not present a significant practical challenge. Well-tested symplectic integrators, predominantly based on the \citet{1991AJ....102.1528W} mapping, are widely available \citep{1998AJ....116.2067D, 1999MNRAS.304..793C, 2019MNRAS.485.5490R, 2019MNRAS.489.4632R}, bringing precise modeling of the distant solar system's long-term evolution within reach of virtually any modern Ghz-grade machine. Nevertheless, such numerical experiments can only solve for the emergent dynamics, not illuminate their theoretical basis. In other words, accurate realizations of orbital evolution can only expose \textit{what} the SDOs do, not \textit{why} they do it. Understanding the latter question requires a simplified analytic model. Here we develop such a model for the distant scattered disk with an eye towards quantifying its underlying dynamical structure and elucidating the processes that drive orbital diffusion, from analytic grounds. We begin by sketching out the statement of the problem. 

\begin{figure*}[t]
\centering
\includegraphics[width=0.8\textwidth]{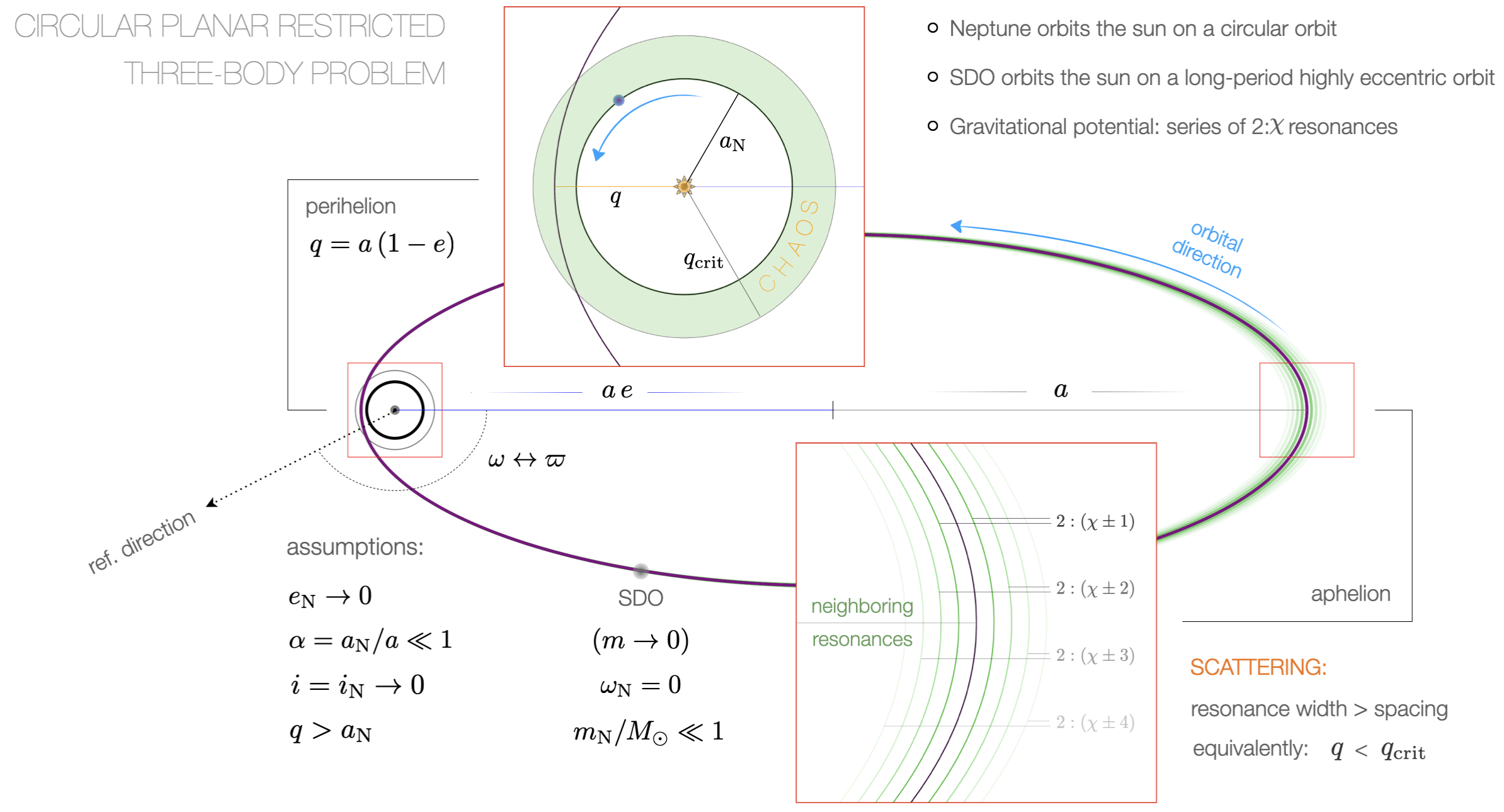}
\caption{Scattered disk dynamics modeled as a circular planar restricted three-body problem. A long-period scattered disk object (SDO) is envisioned to orbit the sun on a highly eccentric orbit with a perihelion distance that exceeds Neptune’s semi-major axis by a small margin. The SDO orbit is shown as a purple ellipse in the digram. Mutual inclination between Neptune and the SDO, as well as perturbations from other planets are neglected. The SDO is modeled as a test-particle. A quadrupole-level spherical harmonic expansion of Neptune’s gravitational potential illuminates that SDO evolution is facilitated by an infinite chain of Neptune's exterior $2:\chi$ resonances. Within the stochastic layer, dynamics of the test particle are primarily driven by the nearest $2:\chi$ resonance (where $\chi$ is an integer approximation to $2\,(a/a_{\rm{p}})^{3/2}$), while its chaotic evolution is facilitated by interactions with neighboring resonances. On the diagram, the nominal locations of $2:\chi\pm1$, $2:\chi\pm2$, $2:\chi\pm3$, and $2:\chi\pm4$ resonances adjacent to the SDO orbit are shown as green ellipses. As we discuss in the text, for the problem at hand, the resonance overlap criterion can be recast as a critical perihelion distance, $q_{\rm{crit}}$, below which chaotic evolution ensues.}
\label{Fig:SETUP}
\end{figure*}

\paragraph{Statement of the Problem} The principal goal of the calculation we aim to carry out is easy to summarize: we wish to develop a simple theory for the long term behavior of highly eccentric, long-period minor bodies, subject to perturbations from Neptune. In other words, our goal is to solve the circular restricted three-body problem in a regime where the test particle possesses an orbital period much larger than that of the perturber, but still experiences material interactions with the planet, owing to the closeness of the perihelion distance to the planet's semi-major axis. The geometric setup of the problem is summarized in Figure \ref{Fig:SETUP}.

The circular restricted three-body problem is by no means a new problem, and the relevant literature spans centuries. Nevertheless, the vast majority of perturbation theory devoted to understanding the relevant dynamics is unsuitable for the problem at hand. Classical expansions of the planetary disturbing function (\citealt{1995CeMDA..62..193L, 2000Icar..147..129E} and the references therein) treat eccentricities and inclinations as small parameters, developing the governing Hamiltonian as a power-series in $e$ and $i$, while placing no constraints on the semi-major axis ratio with the exception of the formal requirement that the orbits do not cross. Conversely, the scattered disk is characterized by large (even near-unity) eccentricities, placing it outside of the domain of applicability of standard models.

As a means to circumvent the limitations of classical methods, various authors have reframed long-term evolution of the scattered disk as mapping problem. That is, rather than attempting to formulate a conventional perturbation theory, \citet{Malyshkin1999, 2004AJ....128.1418P, 2013Icar..222...20F, 2020PASP..132l4401K} envisioned the dynamics as a process wherein the test particle executes unperturbed Keplerian motion, with the exception of the perihelion, where it receives an energy kick of some magnitude that generally depends on the planetary mass and semi-major axis, as well as the particle's perihelion distance. Intuitive in its own right, the essence of this approach lies in the so-called \textit{Kepler Map} (see \citealt{2011NewA...16...94S} for a review). It is worth noting that this mapping was first derived by \citet{1986PhLA..117..328P}, and has since become an important tool for understanding a variety of physical phenomena, including those beyond the realm of dynamical astronomy \citep{Chirikov1989, 1987JPhB...20.5051G, 1988IJQE...24.1420C, 1988JPhB...21L.527J, 1994PhRvA..50..575S, 1998PhLA..241...53S}.

Despite the successes of the mapping approach in modeling chaotic motion at a vastly reduced computational cost, a full understanding of scattered disk dynamics remains incomplete. In particular, the crucial questions of which resonances underly the stochastic layer, and how the scattering process connects to a perturbative description of particle motion, remain to be elucidated. To address this issue, in this work we take an approach that is similar in spirit -- but not in detail -- to classical perturbation theories. More specifically, we adopt a description of the disturbing function as an infinite series developed in terms of a small parameter, which we take to be the semi-major axis ratio, $\alpha=\an/a$, rather than the eccentricity \citep{1962AJ.....67..300K,Laskar2010DISTFUNCT, 2013MNRAS.435.2187M}. As we show below, this Kaula-type expansion attractively lends itself to a simplified description of SDO dynamics and naturally illuminates the relationship between Neptune's exterior mean motion resonances and the scattering process.

The remainder of this paper is organized as follows. In section \ref{sec:anmodel}, we outline the basis of our analytical theory. The results are presented in section \ref{sec:results}. Particularly, in section \ref{sec:intmod}, we derive a simplified Hamiltonian model of particle motion. We formulate the stability boundary of the scattered disk in terms of a critical perihelion distance in section \ref{sec:chirikov}. We validate our analytic results with $N-$body simulations in section \ref{sec:numsim}. In section \ref{sec:standard}, we highlight the connection between our model and the Chirikov Standard Map, thus outlining the equivalence between our perturbative approach and scattering viewpoint. Finally, in section \ref{sec:analytic} we derive analytic estimates of the Lyapunov time and the semi-major axis diffusion coefficient within the scattered disk. We summarize and discuss our results in section \ref{sec:discuss}.

\section{Perturbation Theory} \label{sec:anmodel}
As a starting point in our calculation, let us outline our basic assumptions. First and foremost, we treat the SDO as a massless test particle, and assume that its orbital period exceeds that of Neptune by a large margin (i.e., $\alpha\ll1$). Second, we assume that Neptune's eccentricity, $\en$, is sufficiently small to be negligible for our purposes. Third, we neglect all inclination-node dynamics, reducing the problem to a common plane (Figure \ref{Fig:SETUP}). 

\subsection{The Disturbing Function}
A general spherical harmonic expansion of the planetary disturbing function for the co-planar three-body problem is presented in \citet{2013MNRAS.435.2187M}. Employing the usual notation of celestial mechanics, the disturbing function is expressed as a quadruple infinite series:
\begin{align}
\R&=\frac{\G\,\mn}{a}\, \sum_{\ell=2}^{\infty} \, \sum_{m=m_{\rm{min}},\,2}^{\ell} \, \sum_{j' = -\infty}^{\infty} \, \sum_{j = -\infty}^{\infty} \zeta_m\, c_{\ell\,m}^2\,\mathcal{M}_\ell \nonumber \\
&\times \alpha^\ell\, X_{j'}^{\ell,m} (\en) \,X_{j}^{-(\ell+1),m} (e)\, \nonumber \\
&\times \cos\big(j'\,\Mn -j\,M+m(\omega-\omega_{\rm{N}})\big).
\label{R}
\end{align}
In the above expression, the unmarked variables ($a,M,e,\omega$) refer to the orbital elements of the SDO, while those with the subscript $\rm{N}$ correspond to Neptune\footnote{Technically, in equation (\ref{R}), Neptune's mass $\mn$ should be the reduced mass $\mu=\mn\,\Msun/(\mn+\Msun)\approx\mn$, but our choice to replace $\mu$ with $\mn$ is of no practical consequence.}. Note that for the planar problem, the distinction between the longitude and argument of perihelion vanishes, such that $\omega$ simply corresponds to the azimuthal orientation of the apsidal line (see Figure \ref{Fig:SETUP}). We will remark on the dimensionless constants $\zeta, c, \mathcal{M}$, as well as Hansen coefficients $X$ in greater detail below.

In equation (\ref{R}), the index $\ell$ informs the \textit{degree} of the spherical harmonic expansion. Because we are specifically interested in the $\alpha\ll1$ limit, our needs are sufficed by truncating the expansion at quadrupolar level, corresponding to $\ell_{\rm{max}}=2$. This removes the first sum of the series completely, as well as any dependence on the mass-factor $\mathcal{M}$ because $\mathcal{M}_2=1$ for all mass ratios \citep{2013MNRAS.435.2187M}. On a quantitative level, however, this assumption restricts the applicability of our model to long-period (e.g., $a\gtrsim400\,$AU) orbits.

Beyond the first sum, truncation of the series at $\ell=2$ sets manageable bounds of the \textit{order} of the expansion, $m$, such that the second sum runs from $m_{\rm{min}}=0$ to $m_{\rm{max}}=2$. Carrying on, the third sum can be eliminated fully, thanks to an important property of the Hansen coefficient $X_{j'}^{l,m}$. Specifically, it is possible to demonstrate that to leading order in eccentricity, $X_{j'}^{l,m}(\en)=\order(\en^{|m-j'|})$ \citep{1981CeMec..25..101H}, meaning that within the context of our adopted limit of $\en\rightarrow0$, all terms with $m\ne j'$ vanish. The dependence on the first Hansen coefficient in the series is further trivialized by the fact that $X_0^{2,0}(0) = X_1^{2,1}(0) = X_2^{2,2}(0) = 1$. In addition, because apsidal orientation is ill-defined at null eccentricity, we can set $\omega_{\rm{N}}=0$ without loss of generality.

A final simplification comes from the functional form of the constant $c_{\ell\,m}^2$. Written explicitly in terms of spherical harmonics $Y_{\ell,m}$, we have $c_{2\,m}^2=(8\,\pi/5)\,(Y_{2,m}(\pi/2,0))^2$. Crucially, $Y_{2,1}(\pi/2,0)=0$, meaning that all terms of order unity have zero amplitude, and the expansion only contains harmonics with $m=0$ and $m=2$. Writing out all remaining constants explicitly, we have $c_{2\,0}^2 = 1/2$, $c_{2\,2}^2 = 3/4$, $\zeta_0=1/2$, and $\zeta_2=1$. 

With the approximation scheme outlined above, the full quadrupole-level expression for the disturbing function takes the form:
\begin{align}
\R_{\rm{q}}&=\frac{\G\,\mn}{4\,a}\, \alpha^2  \sum_{j = -\infty}^{\infty} \bigg[\underbrace{X_{j}^{-3,0}\, \cos\big(j\,M\big)}_{m\,=\,0} \nonumber \\
&+ \underbrace{3\,X_{j}^{-3,2}\, \cos\big(j\,M- 2(\Mn-\omega)\big)}_{m\,=\,2} \bigg],
\label{Rq}
\end{align}
where the dependence of $X$ on $e$ is implied. Let us now classify these zeroth and second order (in $m$) terms, according to the dynamics they govern.

\subsection{$m=0$: short-periodic and secular terms}
Ignoring the pre-factor in equation (\ref{Rq}), let us begin by examining the first set of harmonics. Noting the general property of Hansen coefficients $X_{j}^{-(\ell+1),m}=X_{-j}^{-(\ell+1),-m}$, it is convenient to write the leading sum as:
\begin{align}
\sum_{j = -\infty}^{\infty} X_{j}^{-3,0}\, \cos\big(j\,M\big) &= (1-e^2)^{-3/2} \nonumber \\
&+2\,\sum_{j = 1}^{\infty} X_{j}^{-3,0} \cos\big(j\,M\big),
\label{sumexp}
\end{align}
where we have taken advantage of the fact that $X_{0}^{-3,0}$ can be evaluated in closed form \citep{1981CeMec..25..101H}.

The first member of the RHS of equation (\ref{sumexp}) is a pure secular term that governs the apsidal precession of an SDO due to the phase-averaged potential of Neptune. The remainder of the RHS is the epitome of short-periodic terms i.e., harmonics that average out on the orbital timescale and contribute virtually nothing to long-term orbital evolution. As is well known, these rapidly oscillating terms can be removed from the Hamiltonian all-together through a near-identity variable transformation, which essentially corresponds to a change from osculating to orbit-averaged (mean) orbital elements (see e.g., Ch. 2 of \citealt{Morbybook}). Thus, we are justified in dropping them from the expression.

Although marginally illuminating, our discussion of $m=0$ perturbations is neither interesting nor new. That is to say, at zeroth order in $m$, the quadrupole-level disturbing function contains no terms that can explain the underlying chaotic structure of the scattered disk. Therefore, scattering dynamics must arise at $m=2$ order, which we examine next. 

\subsection{$m=2$: resonant terms}
The functional form of the $\ell=2,m=2$ harmonic is easy to interpret: the critical argument
\begin{align}
\varphi=j\,M- 2(\Mn-\omega))=j(\lambda-\varpi)-2(\lambda_{\rm{N}}-\varpi)
\label{varphi}
\end{align}
governs the exterior $2:j$ mean-motion resonance with Neptune. Correspondingly, the functional form of equation (\ref{Rq}) indicates that the underlying dynamical structure of the distant scattered disk is nothing more than an infinite sequence of $2:j$ resonances. In fact, to the extent that the quadrupole-level expansion of $\R$ is an accurate representation of the planetary potential, $2:j$ resonances \textit{must} drive the scattering process, since no other harmonics exist in the expansion.

This notion immediately suggests that the stability boundary of the scattered disk (which separates the chaotic and regular dynamics) can be understood within the context of the \citet{Chirikov1979} resonance overlap criterion. We will examine this suspicion more closely below. For the time being, however, we will limit ourselves to simply writing down the model Hamiltonian for the SDO. Dropping short-periodic terms as described above, we have:
\begin{align}
&\Ham=-\frac{\G\,\Msun}{2\,a} - \frac{1}{4} \frac{\G\,\mn}{a}\, \alpha^2 \, (1-e^2)^{-3/2} +\mathcal{T} \nonumber \\
&- \frac{3}{4} \frac{\G\,\mn}{a}\, \alpha^2  \sum_{j = -\infty}^{\infty} X_{j}^{-3,2} \cos\big(j\,M- 2(n_{\rm{N}}\,t-\omega)\big),
\label{Hammy}
\end{align}
where in anticipation of canonical transformations that will follow, we have replaced $\Mn$ with $n_{\rm{N}}\,t$ and introduced a dummy action, $\mathcal{T}$, conjugate to time, in order to keep $\Ham$ formally autonomous.

\subsection{Computation of Hansen Coefficients $X_{j}^{-3,2}$}
The final piece that is needed to complete the specification of our framework is the evaluation of the integral-defined functions $X_{j}^{-3,2}$. Since their introduction by \citet{Hansen1885}, these coefficients have been earnestly studied in the literature (see e.g., \citealt{1970ceme.book.....H,1981CeMec..25..101H, 2008CeMDA.100..287S}), and the general consensus holds that closed-form expressions for coefficients with $j\neq0$ do not exist. Nevertheless, \citet{Sadov2006} has demonstrated that in the double limit of $e\rightarrow1^{-}$ \textit{and} $j\rightarrow\infty$ the specific coefficient $X_{j}^{-3,2}$ (which \citealt{Sadov2006} calls a Chernousko function with index $j-2$) approaches the asymptotic form:
\begin{align}
X_{j}^{-3,2}\,\xrightarrow[j\rightarrow\infty]{e\rightarrow1^{-}}\, -\frac{4}{9}(j-2).
\label{Sadov}
\end{align}

Even the most eccentric scattered disk objects within the current census of TNOs are insufficiently close to $e$ of unity for equation (\ref{Sadov}) to apply. Similarly, a series approximation of $X_{j}^{-3,2}$ in terms of $\sqrt{1-e^2}$ (see \citealt{2008CeMDA.100..287S}) does not converge rapidly enough to be quantitatively useful. Nevertheless, the quasi-linear scaling of $X_{j}^{-3,2}$ with $j$ is intriguing, and through numerical evaluation we have found that the relationship $X_{j}^{-3,2}\propto j$ holds with a surprising degree of accuracy along contours of constant $q=a\,(1-e)=(j/2)^{2/3}\,a_{\rm{N}} \,(1-e)$. Taking advantage of this, the slope of the linear relationship can be expressed as sole function of $q/\an$, and we have found that a simple Gaussian-like parameterization achieves satisfactory precision:
\begin{align}
X_{j}^{-3,2} \approx \frac{2\,j}{5}\exp\bigg[ -\bigg(\frac{q}{a_{\rm{N}}} \bigg)^2\, \bigg].
\label{param}
\end{align}
In fact, applied specifically to the observationally relevant $q\in(30,50)\,$AU range, equation (\ref{param}) agrees with direct evaluation of Hansen coefficients with $j>10$ down to a few percent. We will elaborate on the calculational advantage of evaluating $X_{j}^{-3,2}$ along a locus of constant perihelion further below.

It is interesting to compare the form of equation (\ref{param}) to expression (B5) of \citet{2013MNRAS.435.2187M}, which is based on an asymptotic expansion for the overlap integral representing the energy exchanged during one outer orbit (see also equations 3.55 and 3.73 of \citealt{Mardling2008Chaos}). In particular, the exponential decay of the low and high-eccentricity tails reflects the fact that an exponentially small amount of (specific) energy is exchanged between the orbits when the angular frequency of the test particle at perihelion is significantly different to the orbital frequency of Neptune. Conversely, significant energy is exchanged when these frequencies are similar. In fact, they are the same when $q/\an= (1+e)^{1/3}$ which for $e\sim 1$ corresponds to $q\approx38\,$AU. Thus, even before examining the onset of instability from the vantage point of the Chirikov criterion, we may intuitively expect the semi-major axis dependence of the perihelion stability boundary to be relatively shallow. 

\section{Results} \label{sec:results}
In order to evaluate the stability boundary of the scattered disk established by $2:j$ resonances, we must estimate the critical value of $q$ as a function of $a$, at which neighboring resonances overlap. Accordingly, we now project the separatrixes of the individual resonances onto the $q-a$ plane. As is usual for calculations of this type, the first step is to write down an integrable pendulum-like Hamiltonian for an isolated $2:\chi$ resonance, where $\chi$ is an integer nearest to $2\,(a/a_{\rm{p}})^{3/2}$. 

\subsection{An Integrable Model for an Isolated $2:\chi$ Resonance} \label{sec:intmod}
Because the resonance width is expected to be small compared to the SDO semi-major axis, a conventional approach to circumventing the inverse semi-major axis dependence of the Keplerian term in equation (\ref{Hammy}) is to Taylor-expand it around the nominal resonance location. Accordingly, in terms of conventional Delaunay variables $L=\sqrt{\G\,\Msun\,a}, l=M, G=L\,\sqrt{1-e^2}, g=\omega$ (see Ch. 2 of \citealt{MD99}), we have:
\begin{align}
&-\frac{1}{2}\bigg(\frac{\G\,\Msun}{L} \bigg)^2 \approx -\frac{1}{2}\bigg(\frac{\G\,\Msun}{[L]} \bigg)^2 + \bigg(\frac{\G\,\Msun}{[L]} \bigg)^2 \bigg( \frac{\delta L}{[L]} \bigg) \nonumber \\
& - \frac{3}{2} \bigg(\frac{\G\,\Msun}{[L]} \bigg)^2 \bigg( \frac{\delta L}{[L]} \bigg)^2 =[n]\,\bigg(\delta L - \frac{3}{2}\frac{\delta L^2}{[L]} \bigg)+\dots ,
\label{Hkeptaylor}
\end{align}
where $[L]=\sqrt{\G\,\M\,(\chi/2)^{2/3}\,\an}$ is the nominal action and $[n]=(2/\chi)\,n_{\rm{N}}$ is the mean motion at the center of the $2:\chi$ resonance. At this stage, it is convenient to adopt $\delta L=L-[L]$ as the action instead of $L$ itself, keeping in mind that translation of an action by a constant is always canonical. 

Let us now define a change of variables through a type-2 generating function:
\begin{align}
\mathcal{F}_2=(\underbrace{\chi\,l/2-(n_{\rm{N}}\,t-g)}_{\phi})\,\Phi+(\underbrace{l}_{\psi})\,\Psi+(\underbrace{t}_{\xi})\,\Xi.
\label{Hkeptaylor}
\end{align}
The actions conjugate to the new angles $\phi$, $\psi$, and $\xi$ are defined by the usual transformation equations:
\begin{align}
&\delta L=\frac{\partial\,\mathcal{F}_2}{\partial\,l}=\frac{\chi}{2}\,\Phi+\Psi \nonumber \\
&G=\frac{\partial\,\mathcal{F}_2}{\partial\,g}=\Phi \nonumber \\
&\mathcal{T}=\frac{\partial\,\mathcal{F}_2}{\partial\,t}=\Xi-n_{\rm{N}}\,\Phi.
\label{transform}
\end{align}

With the preliminaries (\ref{Hkeptaylor}) and (\ref{transform}) delineated, we are now in a position to write down an idealized Hamiltonian, $\Ham_{\chi}$, for each isolated resonance. Neglecting the unimportant $m=0$ secular term in equation (\ref{Hammy}) and retaining only the principal harmonic, we have:
\begin{align}
\Ham_{\chi}&=-\frac{3}{4}\frac{n_{\rm{N}}\,\chi}{[L]} \, \Phi^2-\frac{3\,n_{\rm{N}}}{[L]}\,\Psi\,\Phi \nonumber \\
&- \frac{3}{4} \frac{\G\,\mn}{a}\, \bigg(\frac{a_{\rm{N}}}{a} \bigg)^2\,X_{\chi}^{-3,2}\,\cos\big(2\,\phi \big) \nonumber \\
&+\underbrace{\Xi + \frac{2\,n_{\rm{N}}}{[\chi]}\,\Psi-\frac{3\,n_{\rm{N}}}{\chi\,[L]}\,\Psi^2 -\frac{1}{2}\bigg(\frac{\G\,\Msun}{[L]} \bigg)^2}_{\rm{const.}}
\label{Hchi}
\end{align}
Notice that upon switching to variables (\ref{transform}), the linear term in $\Phi$ arising from the $[n]\,\delta L$ term in equation (\ref{Hkeptaylor}) is exactly cancelled by the $-n_{\rm{N}}\,\Phi$ term that ensues from the dummy action $\mathcal{T}$, owing to the fact that $(\chi/2)\,[n]=n_{\rm{N}}$.

$\Ham_{\chi}$ is now independent of the angles $\psi$ and $\xi$, and the conjugate actions $\Psi$ and $\Xi$ are integrals of motion. Accordingly, all terms on the third line of equation (\ref{Hchi}) are constant and can simply be dropped from the Hamiltonian. Moreover, the linear action term (proportional to $\Phi\,\Psi$) can be absorbed into the leading term by adding $(3\,n_{\rm{N}}\Psi^2)/(\chi\,[L])$ to $\Ham_{\chi}$ and completing the square, such that the nonlinear action term becomes proportional to $\tilde{\Phi}^2=(\Phi-2\Psi/\chi)^2$. Then, adopting $\tilde{\Phi}$ as the new action conjugate to $\phi$ (again, by canonical translation) and substituting parameterization (\ref{param}) for the Hansen coefficient, we obtain the Hamiltonian of a mathematical pendulum:
\begin{align}
\Ham_{\chi}=&-\underbrace{\frac{3}{\an^2}\,\bigg( \frac{\chi}{2} \bigg)^{2/3}}_{\beta}\,\frac{\tilde{\Phi}^2}{2}\nonumber \\
&-\underbrace{\frac{6}{5}\frac{G\,\mn}{\an\,\chi}\exp\bigg[-\bigg(\frac{q}{\an}\bigg)^2\,\bigg]}_{\gamma}\,\cos(2\,\phi).
\label{Hchi2}
\end{align}

\begin{figure*}[t]
\centering
\includegraphics[width=\textwidth]{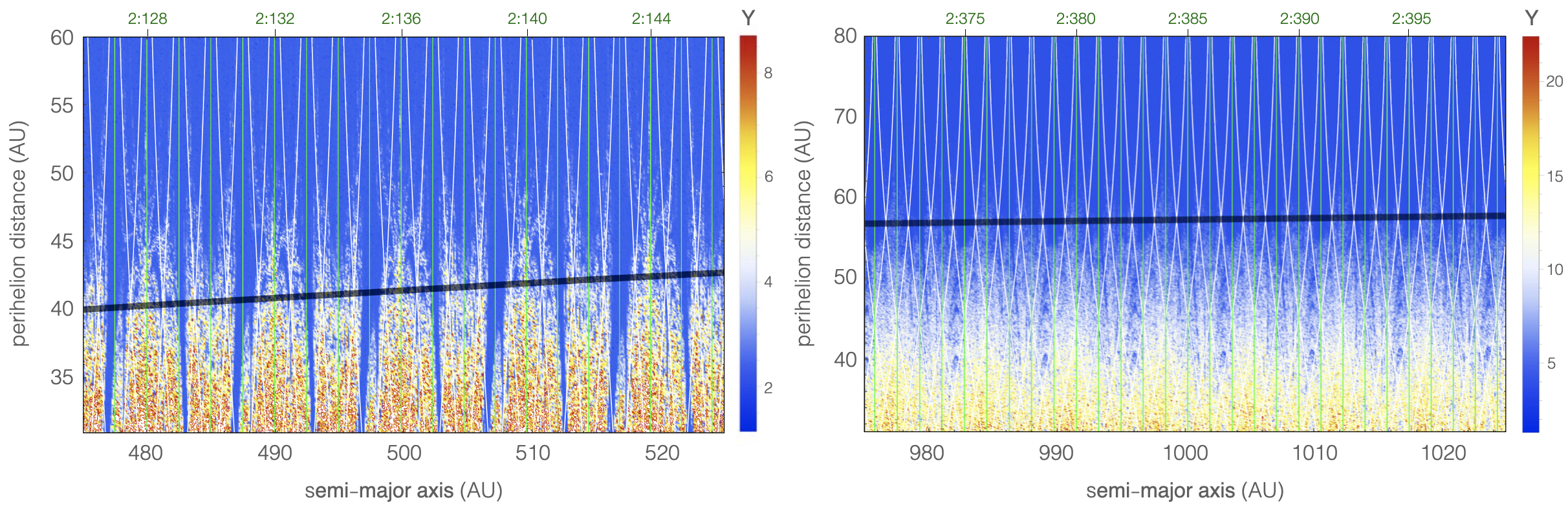}
\caption{Chaos map of the distant scattered disk, modeled within the framework of the circular planar restricted three-body problem. A heat map of the MEGNO chaos indicator, $Y$, is shown on the semi-major axis vs. perihelion plane. Blue regions of the diagram depict initial conditions that lead to regular motion, whereas yellow and red regions correspond to chaotic dynamics. Within the chaotic layer, the Lyapunov time of the SDO approaches the orbital period. The analytic threshold for chaotic motion ($q_{\rm{crit}}$, given by equation \ref{qcrit}) is shown with a thick black line. The nominal locations and widths of individual $2:\chi$ mean motion resonances are shown with thin green and white lines, respectively.}
\label{Fig:MEGNO}
\end{figure*}

It is worth noting that in the well-studied case of low-$e$ mean motion resonance dynamics \citep{MD99,Morbybook} the oscillation period of the resonant angle exceeds the orbital period by a large margin. This is not the case for the SDO scattering problem at hand: in the $q\sim \an$ regime, the ratio of the libration frequency to mean motion is given by $\sqrt{\beta\,\gamma}/n\sim (a/\an)^{5/4}\,\sqrt{\mn/\M}\sim1/4-1/2$. Thus, the orbital frequency exceeds the libration frequency only by a factor of a few.

The expression for the resonance width of a mathematical pendulum is well known: $\Delta\tilde{\Phi}=4\sqrt{\gamma/\beta}$ (Ch. 4 of \citealt{Morbybook}). It is important to understand, however, that this width -- expressed in terms of the canonical action $\tilde{\Phi}$ -- is ultimately related to the SDO's eccentricity. To relate this quantity to the resonance width in terms of the semi-major axis, we use the conservation of $\Psi$. In fact, it is straightforward to demonstrate that the conservation of $\Psi$ is nothing more than a re-statement of the conservation of the Tisserand parameter (Ch. 8 of \citealt{MD99,2013A&A...556A..28B}). Moreover, it is easy to show that for long-period and highly eccentric orbits, the conservation of the Tisserand parameter is equivalent to a conservation of the perihelion distance. A more detailed discussion of the physical meaning of the conservation of $\Psi$, and how it relates to other quasi-integrals of motion of the circular restricted three-body problem is presented in the Appendix.

\subsection{The Chirikov Criterion} \label{sec:chirikov}

From equation (\ref{transform}), it follows that $\Delta\delta L=\chi\,\Delta\tilde{\Phi}/2$. Direct substitution therefore yields: 
\begin{align}
\Delta a = 4\,a_{\rm{N}}\,\sqrt{\frac{2\,\chi\,\mn}{5\,\Msun}} \, \exp\bigg[-\bigg(\frac{q}{2\,\an}\bigg)^2\,\bigg].
\label{deltaa}
\end{align}
We note that while we arrived at this expression from the Hamiltonian formalism, an alternative approach would have been to start with the disturbing function (\ref{Rq}), write down Lagrange's equations of motion, and proceed to derive a pendulum-like equation of motion for the critical argument, $\phi$. Indeed, both approaches yield equivalent results (see e.g., Ch. 8.6. of \citealt{MD99}, section 2.3 of \citealt{2013MNRAS.435.2187M}; see also \citealt{1980AJ.....85.1122W, Mardling2008Chaos}). 

As stipulated by \citet{1959SPhD....4..390C,Chirikov1979}, the width of the resonance, $\Delta a$ should be compared with the distance between adjacent resonances, $\delta a = ([a]_{\chi+1}-[a]_{\chi-1})$. In the limit of large $\chi$, it is straightforward to show that $\delta a \approx (2\,\an/3)\,(2/\chi)^{1/3}$. The degree of resonance overlap is characterized by the ratio of $\Delta a$ and $\delta a/2$ (recall that neighboring resonances also widen in an equivalent way). Expressing this number in terms of SDO semi-major axis instead of $\chi$, we have:
\begin{align}
\frac{\Delta a}{\delta a} = \frac{24}{\sqrt{5}}\,\bigg(\frac{a}{\an}\bigg)^{5/4}\,\sqrt{\frac{\mn}{\Msun} }\,\exp\bigg[-\bigg(\frac{q}{2\,\an}\bigg)^2\,\bigg].
\label{K}
\end{align}
This result demonstrates an intriguing trend: along a locus of constant perihelion distance, the degree of overlap \textit{grows} with increasing particle semi-major axis. As importantly, we can set the overlap number equal to the critical value of unity ($\Delta a/\delta a\rightarrow1$), and invert this relation:
\begin{align}
q_{\rm{crit}}=\an\,\sqrt{\ln\bigg( \frac{24^2}{5}\,\frac{\mn}{\Msun }\,\bigg(\frac{a}{\an}\bigg)^{5/2}\bigg)}. 
\label{qcrit}
\end{align}
This expression yields a critical perihelion distance, $q_{\rm{crit}}$, as a function of semi-major axis, below which chaotic diffusion is expected to ensue. 

\subsection{Numerical Validation}  \label{sec:numsim}
The analytic calculations outlined above yield a compact result for the chaotic threshold of the scattered disk. This result is a key prediction of our theory that can be tested with numerical experiments in a straightforward manner. Here, we carry out this examination as a sequence of two sets of $N-$body simulations employing distinct levels of complexity. More specifically, our first task is to compare our expression with a chaos map generated within the context of an identical physical configuration -- the circular planar restricted three-body problem. This Sun-Neptune-SDO setup provides the closest point of comparison between our analytic theory and numerical calculations, and is equivalent to lifting the assumptions employed in reducing the complexity of the disturbing function (\ref{R}).


A customary way to map out the boundaries between chaotic and regular motion is to compute the system's Lyapunov coefficient, $\Lambda$ (or its siblings), on a plane of initial conditions. Here, we follow this conventional approach, substituting the Lyapunov coefficient for the more-rapidly-convergent MEGNO chaos indicator, $Y$ \citep{MEGNO}. We carried out these simulations using the \texttt{REBOUND} gravitational dynamics software package \citep{2019MNRAS.485.5490R,2019MNRAS.489.4632R}, employing the \texttt{whfast} integration algorithm with an initial time-step of $\delta t=63$ code units\footnote{The code uses units where the gravitational constant $\G$ is set to unity, such that in a unit system that employs solar masses and astronomical units, this time-step corresponds to approximately 10 years, or equivalently, $6\%$ of Neptune's orbital period.}. We generated two such maps, with $a\sim500\,$AU and $a\sim1000\,$AU. Each integration spanned $\Delta t = 0.1\,$Myr for $a\sim500\,$AU runs but was increased to $\Delta t = 0.3\,$Myr for $a\sim1000\,$AU runs to accommodate the longer orbital period. The resolution of our grid of initial conditions in SDO perihelion distance and semi-major axis was set to $\delta q = \delta a = 0.1\,$AU. Neptune's eccentricity remained at $e_{\rm{N}}=0$ throughout the integrations. Additionally, all starting orbital angles were set to null values, with the exception of the SDO mean anomaly, which was initialized at $M=\pi$ (aphelion). 

The left and right panels of Figure \ref{Fig:MEGNO} show MEGNO maps centered around a SDO semi-major axes of $a=500$ and $1000\,$AU, respectively. On the same plane, we mark the locations of individual $2:j$ resonances with green lines and project their widths according to equation (\ref{deltaa}) with white curves. The critical perihelion distance, corresponding to marginal overlap given by equation (\ref{qcrit}), is shown with a thick black line. As the color-bar indicates, blue regions of the plot (where $Y\sim2$) correspond to regular motion while initial conditions depicted with red and yellow points (where $Y\sim\Lambda\,\Delta t/2$) indicate chaotic SDO dynamics. As a check on our simulations, we recomputed portions of the shown MEGNO maps with a different choice of integration algorithm (\texttt{IAS15}) and longer timespan ($3\times\Delta t$) and got equivalent results.

Overall, the analytic criterion (\ref{qcrit}) provides a satisfactory approximation for the boundary between regular particle motion and large-scale chaos. Nevertheless, we remark that this threshold is inexact, and fine structure, including that arising from higher-order resonances, causes equation (\ref{qcrit}) to underestimate the critical value of $q$ at some values of $a$ while overestimating it at others. To elaborate on this further, the fact that regular regions exist at $q<q_{\rm{crit}}$ may in part be attributed to the fact that regular islands exist within the chaotic sea even if there is substantial overlap. The existence of chaotic regions for $q>q_{\rm{crit}}$, however, likely illuminates the limitations of our analytic model. To this end, it is likely that a more detailed resonance overlap criterion that also accounts for octupole-level resonances could generate better agreement. Note further that the agreement between $N-$body simulations and our theory is somewhat better for $a=1000\,$AU than for $a=500\,$AU. This is not surprising, given that the assumptions of our model are better satisfied for increasingly long-period orbits.  


\begin{figure}[t]
\centering
\includegraphics[width=\columnwidth]{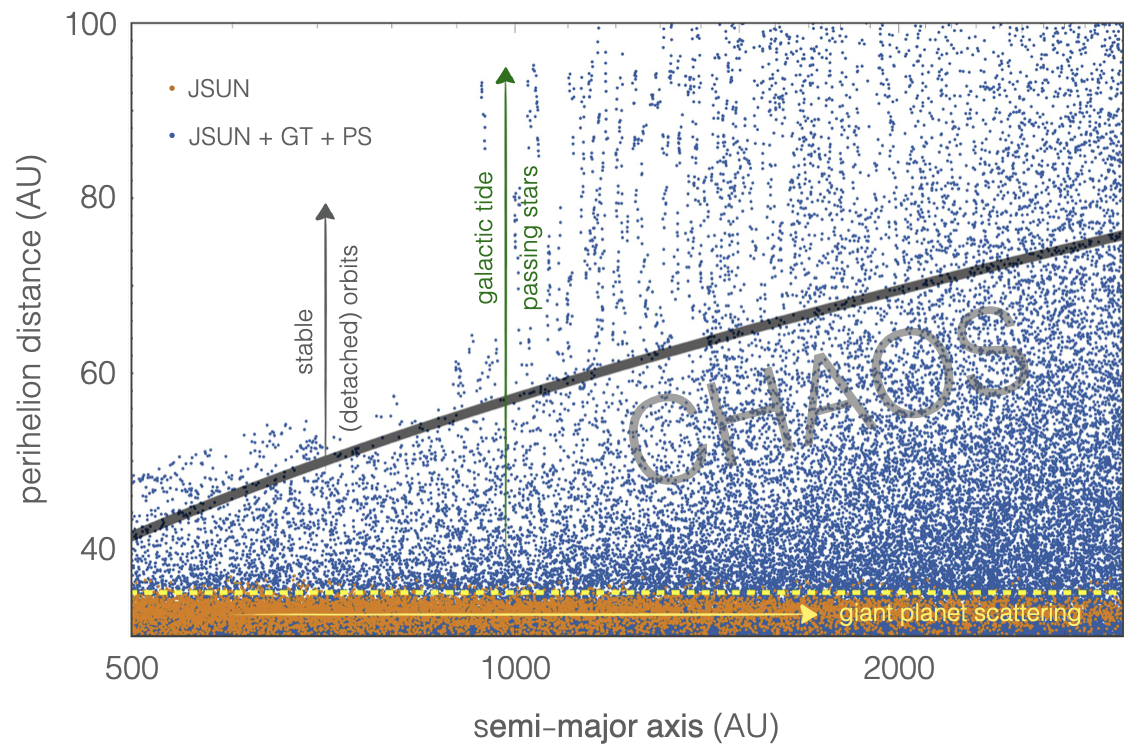}
\caption{Detailed models of the distant scattered disk. Orange points depict a model of an evolved scattered disk that is created exclusively by giant planet scattering, accounting for early migration of Neptune through the solar system. The purple points show a model scattered disk that is affected by giant planets as well as the galactic tide and passing stars. An inclination cut of $i<40\deg$ was applied to both models. The analytic threshold for chaos is shown with a thick black curve, as in Figure \ref{Fig:MEGNO}. While the resonance overlap criterion marks the boundary between regular and stochastic dynamics, it should not be interpreted as the boundary of the scattered disk itself. In an idealized scenario that only includes giant planet scattering, the near-conservation of the Tisserand parameter prevents SDOs from filling the entirety of the chaotic domain. In a more realistic model that also accounts for extrinsic effects, Galactic perturbations can raise and lower SDO perihelia across the chaotic threshold.}
\label{Fig:NUMSIM}
\end{figure}

With our analytic expression for the chaotic boundary verified through numerical experimentation, we now consider how this threshold for orbital stability compares with detailed models of the formation and evolution of the scattered disk. To this end, we reference the published simulation suite of \citet{Nesvorny2017}, where the genesis of the scattered disk was simulated accounting for the early outward migration of Neptune \citep{2005Natur.435..459T,2011ApJ...738...13B, 2016ApJ...825...94N}, and its long-term fate was self-consistently modeled subject to gravitational forcing from the giant planets as well as (optionally) the Galactic tide and passing stars.

The orbital structure of evolved ($t=4.5\,$Gyr) synthetic models of the distant scattered disk with $i<40\,\deg$ are contrasted against our analytic stability boundary in Figure \ref{Fig:NUMSIM}. More specifically, the results of a simulation where extrinsic effects were omitted are shown with orange points, while a scattered disk that is sculpted by Galactic tides and passing stars in concert with the planets is shown with purple dots. Upon examination, an important conclusion can immediately be drawn: the boundary of the scattered disk (meaning the parameter space occupied by the particles) does \textit{not} uniformly trace its chaotic threshold. That is, in absence of Galactic forcing, long-period particles retain relatively low perihelia with $q\lesssim36\,$AU and do not extend to the edge of the chaotic zone. Conversely, when the effects of the Galactic tide and passing stars are included, the resulting eccentricity modulation can lift the perihelia of SDOs well above the critical value for chaos, especially for $a\gtrsim1000\,$AU orbits.

These results can be understood within the framework of our model as follows. While the $q<q_{\rm{crit}}$ orbital domain is largely chaotic, the long-period SDO dynamics nevertheless approximately obey the conservation of the Tisserand parameter. As shown in the Appendix of this work, preservation of the Tisserand parameter (or analogously the resonant integral of motion $\Psi$ defined in equation \ref{transform}) is equivalent to evolution along a constant-perihelion contour for orbits with $a\gg\an$ and $e\sim1$. This near-conservation of the perihelion distance prevents SDOs from exploring the full range of parameter space spanned by the chaotic sea in simulations that only include planetary forcing. 

The opposite situation ensues in numerical experiments that include the Galactic tide. Under the action of the Galactic tide, all symmetry inherent to the circular restricted three-body problem is broken, allowing significant $q$ variation to take place. Accordingly, at sufficiently long orbital periods, SDOs can be carried to large perihelion distances with no regard for the chaotic boundary facilitated by Neptune. The transition between scattering-dominated dynamics and evolution primarily driven by the Galactic tide is relatively sharp, and occurs at a semi-major axis of $a\gtrsim1000\,$AU. Qualitatively, this shift corresponds to a point where the timescale associated with von Zeipel-Lidov-Kozai type perihelion oscillations facilitated by the Galactic tide becomes markedly shorter than the perihelion precession timescale forced by the giant planets.

The dynamical origins of $q\gtrsim36\,$AU $a\lesssim1000\,$AU objects are considerably more subtle. Notice that unlike their more distant counterparts, these objects follow the stability boundary of the scattered disk relatively well. Owing to comparatively rapid perihelion precession, the perihelia of these objects cannot be affected by the Galactic tide directly. Nevertheless, their lowered eccentricities indicate that they have been materially affected by the Galactic tide \textit{at some point}, implying that they must have attained $a\gtrsim1000\,$AU in the past. Correspondingly, these are objects that initially get scattered onto large heliocentric distances, and after significant Galactic perturbation diffuse back to smaller semi-major axes. As inward semi-major axis diffusion gets terminated at the chaos boundary, parameter space traced by $q_{\rm{crit}}$ gets filled in from the outside. Examination of individual time-series of particles in the simulations confirms this interpretation.





\subsection{Linking the Scattered Disk with the Modulated Pendulum and the Standard Map} \label{sec:standard}
Against the backdrop of the perturbative treatment of the dynamics developed in the preceding sections, it is important to not forget that the more rudimentary -- but somewhat more physically intuitive -- picture of scattered disk dynamics is one wherein perturbations are envisioned as ``kicks" to the orbit that ensue when the SDO passes through perihelion and experiences a gravitational interaction with Neptune \citep{2004AJ....128.1418P, 2013Icar..222...20F}. Accordingly, it is useful to briefly examine the connection between our perturbative framework and this ``mapping" viewpoint.

To begin making the analogy, note that in the limit of large $\chi$, the Hansen coefficients $X_{\chi}^{-3,2}\approx X_{\chi \pm 1}^{-3,2}$. Thus, let us assume that the Hansen coefficients with neighboring indexes are not simply similar, but are in fact, equal to one-another. Under this approximation, we can factorize the Hansen coefficient in equation (\ref{Hammy}), to obtain a simple non-autonomous Hamiltonian that accounts for interactions between the primary $2:\chi$ resonance and its nearest neighbors:
\begin{align}
\Ham_{\chi\pm}&=\beta\,\frac{\tilde{\Phi}^2}{2}-\gamma\,\big( \cos(2\,\phi-l)+\cos(2\,\phi)+\cos(2\,\phi+l)\big)\nonumber \\
&=\beta\,\frac{\tilde{\Phi}^2}{2}-\gamma\,(1+2\cos(n\,t)) \,\cos(2\,\phi).
\label{modpend}
\end{align}
Note that here we have used a trigonometric identity and set $l=M=n\,t$ to arrive at the second line (recall further that $\beta$ and $\gamma$ are defined in equation \ref{Hchi2}). This expression corresponds to the Hamiltonian of a \textit{modulated pendulum}, where the modulation frequency is equal to the SDO's mean-motion (Ch. 4 of \citealt{Morbybook}). Recalling that the mean motion is faster than the libration frequency by a factor of a few, chaotic dynamics that arise within the context of our problem lie squarely outside of the ``adiabatic" domain.

Let us now push our luck, and extend the aforementioned approximation by assuming that \textit{all} Hansen coefficients in the infinite perturbation series are equal. Although seemingly crude, this approximation in fact holds relatively well in practice because the dynamics of any given resonance is most strongly affected by perturbations that are ``nearby" in action space (or equivalently, in frequency space). Indeed, the amplitudes of faraway resonances do not matter much, since the harmonics vary rapidly and the corresponding terms quickly average out (see e.g., \citealt{Wisdom1982} for a discussion). In this limit, we can imagine that the sum in equation (\ref{Hammy}) runs exclusively over the cosines. Thus, employing a Fourier representation of the periodic $\delta$-function, we can write:
\begin{align}
&\sum_{j=-\infty}^{\infty}\cos(j\,l+2\,\phi)=\cos(2\,\phi)\,\sum_{j=-\infty}^{\infty}\cos(j\,l) \nonumber\\
&= \frac{1}{2\,\pi} \cos(2\,\phi) \, \delta_{2\pi/n},
\label{deltacos}
\end{align}
where $\delta_{2\pi/n}$ represents an impulse comb that is applied with the orbital period of the SDO at $l=0$ (perihelion).

Substituting equation (\ref{deltacos}) back into the expression for $\Ham$, we see that when expanded in the vicinity of a $2:\chi$ resonance, Hamiltonian (\ref{Hammy}) takes on the familiar form of a periodically kicked pendulum: 
\begin{align}
\Ham=\beta\,\frac{\tilde{\Phi}^2}{2}- \frac{\gamma}{2\,\pi} \cos(2\,\phi)\, \delta_{2\pi/n}.
\label{kicked}
\end{align}
As is well known, Hamiltonian (\ref{kicked}) generates the \textit{Chirikov Standard Map} -- an emblematic model of chaotic dynamics (e.g., \citealt{LLbook,Chirikov1979}). In fact, the appearance of the Standard Map within the context of this problem acts as the bridge between our analytic framework and the scattering viewpoint. To this end, it is crucial to note that the Kepler Map discussed in section \ref{sec:intro}, is locally identical to the Standard Map, which is governed by the above Hamiltonian \citep{2011NewA...16...94S, 2009IJMPD..18.1903K}. The connection between the perturbative treatment of SDO evolution and a mapping approach to modeling the orbital motion is thus clear.

\subsection{Chaos in the Scattered Disk: Analytic Estimates} \label{sec:analytic}
An important motivation behind making the connections between our perturbative theory of scattered disk dynamics and archetypal models of chaotic motion described above, is that the latter naturally lend themselves to analytic estimates \citep{LLbook}. In this vein, previous work aimed at quantifying Lyapunov times and the action diffusion constants of main belt Asteroids \citep{1996AJ....112.1278H,1997AJ....114.1246M,1998CeMDA..71..243N} and Mercury \citep{Laskar2008,2011ApJ...739...31L, 2015ApJ...799..120B} played an important role in expanding our overall understanding of chaotic small body evolution within the inner solar system. Here we continue this program, and focus on quantifying the Lyapunov time and semi-major axis diffusion coefficient within the scattered disk, from analytic grounds.  

\paragraph{Lyapunov Time} Our estimate the SDO Lyapunov time, $\tau_{\rm{L}}$, follows directly from the analogy with a modulated pendulum equation (\ref{modpend}) made above. To outline the qualitative picture, recall that the resonance width of a mathematical pendulum scales as the square root of the factor that multiplies the harmonic term of the Hamiltonian. Because this factor is time-dependent in equation (\ref{modpend}), however, the separatrix in our problem is not steady, and instead pulsates at the modulation frequency. In the regime of strong resonance overlap -- which we can crudely assume for orbits with $q< q_{\rm{crit}}$ -- a large fraction of the resonant phase-space area is periodically swept by a homoclinic curve, that instills hyperbolicity upon the SDO trajectory with the same frequency (Ch. 9.4 of \citealt{Morbybook}). Therefore, to an order of magnitude, the SDO's Lyapunov time can be interpreted as the modulation period, which in the case of Hamiltonian (\ref{modpend}) is nothing other than the orbital period:
\begin{align}
\tau_{\rm{L}}\sim\Lambda^{-1}\sim\frac{2\,\pi}{n}=\sqrt{\frac{4\,\pi^2\,a^3}{\G\,\Msun}}.
\label{Lyapana}
\end{align}

The fact that the Lyapunov time in the scattered disk is comparable to the orbital period can be understood from intuitive grounds. While macroscopic divergence of neighboring trajectories may require multiple Lyapunov times to ensue (depending on the initial separation of nearby starting conditions), it is important to keep in mind that $\tau_{\rm{L}}$ itself is a measure of decoherence on a microscopic scale. Accordingly, two initially nearby trajectories within the scattered disk will experience perturbations from Neptune at slightly distinct phases, meaning that their separation in phase-space will be amplified on the orbital timescale.

To test this assertion, let us return to Figure \ref{Fig:MEGNO} and examine the values of the MEGNO chaos indicator that ensue within the stochastic layer. At $a=500\,$AU, where the SDO orbital period is approximately $11{,}000$ years, the chaotic domain is characterized by $Y\sim9$. Recalling that $Y\sim2\,\Delta t/\tau_{\rm{L}}$ with $\Delta t=0.1\,$Myr, we thus obtain $\tau_{\rm{L}}\sim2\times10^{4}\,$years -- a value comparable to the orbital period. We have further checked these results with a few traditional calculations of the Lyapunov times through direct integration of the variational equations (for SDOs randomly initialized with $31<q<36$ and $a=500\,$AU; \citealt{2016MNRAS.459.2275R}) and obtained estimates of $\tau_{\rm{L}}$ that were even closer to the orbital period. The MEGNO map at $a=1000\,$AU tells a similar story: with $\Delta t = 0.3\,$Myr and a characteristic $Y\sim20$, we obtain $\tau_{\rm{L}}\sim3\times10^{4}\,$years -- a value very close to the approximately $31{,}000$ year orbital period. 

\paragraph{Diffusion Coefficient} It is well established that within a stochastic system subject to vigorous mixing, the statistical properties of the actions obey the Fokker-Plank equation \citep{1945RvMP...17..323W}. Moreover, if the system is Hamiltonian, it can be shown that the Fokker-Plank equation reduces to the conventional diffusion equation, such that all of the relevant physics is encapsulated in the diffusion coefficient, $\mathcal{D}$. 

In the quasi-linear approximation, the value of $\mathcal{D}$ can be generally estimated as the product of the Lyapunov coefficient and the square of the resonant half-width \citep{Chirikov1979, LLbook}. The physical interpretation of this relation is that the resonant half-width represents a typical stochastic ``step-size" that a trajectory attains over a single decoherence (Lyapunov) time. For the problem at hand, the resonance half width, $\Delta a/2$, follows from equation (\ref{deltaa}), and we have already shown that $\tau_{\rm{L}}$ is well-approximated by the orbital period\footnote{Similar dynamics can arise in the case of first order resonances of a high degree, where a kick received during conjunction can produce changes in action that are comparable with the resonance width \citep{1994A&A...289..972S}}. The semi-major axis diffusion coefficient thus has the form:
\begin{align}
\mathcal{D}_a\sim\frac{\Delta a^2}{4\,\tau_{\rm{L}}}&=\frac{8}{5\,\pi}\frac{\mn\,\sqrt{\G\,\Msun\,\an}}{\Msun} \nonumber \\
&\times \exp\bigg[-\frac{1}{2}\bigg(\frac{q}{\an}\bigg)^2\,\bigg].
\label{Diffcoeff}
\end{align}
Note that this expression is independent of the particle's semi-major axis, and only depends on its perihelion distance. 

As a numerical check on our assumption that $\Delta a/2$ is truly a suitable approximation for a characteristic semi-major kick experienced by an SDO over a single orbital period, we ran 500 single-orbit Sun-Neptune-SDO simulations with $q=35\,$AU, randomized phases, and semi-major axis sampled uniformly in the $a=500\pm5\,$AU range. We then measured the aphelion-to-aphelion variation in particle semi-major axes, and found a mean value of $2.32\,$AU, in good agreement with the results of \citet{2013Icar..222...20F}. This quantity is close to the theoretically predicted value of $\Delta a/2 = 2.26\,$AU, leading us to conclude that equation (\ref{Diffcoeff}) provides an adequate approximation for the semi-major axis diffusion coefficient of long-period scattered disk objects.


\section{Discussion} \label{sec:discuss}

Owing to the unrelenting observational mapping of the trans-Neptunian solar system that has ensued over the last two decades, the orbital structure of the scattered disk continues to come into an ever-shaper focus. Several attempts have been made to describe the stochastic dynamics of this remarkable population of minor bodies. In this vein, $1:j$ resonances have been broadly discussed in the literature as an attractive theoretical explanation for the emergent behavior of actively scattering TNOs \citep{2004AJ....128.1418P, 2018AJ....155..260V, 2019CeMDA.131...39L}. Nevertheless, a complete understanding of the evolution of long-period orbits has remained incomplete.

In this work, we have approached the problem of scattered disk dynamics from a perturbative viewpoint. In particular, we have derived a simple Hamiltonian model for the orbital motion of long-period TNOs, based upon a quadrupole-level expansion of the planetary disturbing function \citep{1962AJ.....67..300K,Laskar2010DISTFUNCT,2013MNRAS.435.2187M}. Our analysis indicates that the scattered disk's dynamical machinery is comprised of a chain of $2:j$ resonances and that their overlap is responsible for driving chaotic motion. To be clear, $1:j$ harmonics are not entirely absent from the dynamical picture, but are smaller than $2:j$ resonances by a factor of $\en$ at quadrupole order, or a factor of $\alpha$ (i.e., appearing at octupole+ order) in the $\en\rightarrow0$ limit. We further demonstrate how our theoretical model can be reduced to the Chirikov Standard Map \citep{Chirikov1979}, illuminating the physical connection between resonant perturbations and the scattering process itself.

Interpreting the intersection point among nonlinear $2:j$ resonances as the dividing line between regular and stochastic motion, we have derived an analytic stability boundary of the distant scattered disk. In practice, this criterion is given by equation (\ref{qcrit}) and translates to a critical perihelion distance below which chaos ensues. For chaotic orbits that satisfy this criterion, we have obtained analytic estimates of Lyapunov time, $\tau_{\rm{L}}$ (equation \ref{Lyapana}), and the semi-major axis diffusion coefficient, $\mathcal{D}_a$ (equation \ref{Diffcoeff}). Importantly, these calculations indicate that within the strongly chaotic domain of the scattered disk, the Lyapunov time approaches the orbital period, while the semi-major axis diffusion coefficient is on the order of Neptune's angular momentum divided by the solar mass. Our analysis further shows that the semi-major axis diffusion rate (or equivalently, the rate of energy diffusion) is insensitive to the semi-major axis itself. Instead, $\mathcal{D}_a$ only depends on the perihelion distance -- a result that is consistent with previous findings \citep{2004AJ....128.1418P, 2013Icar..222...20F}.

Although compact and easy to implement, we caution that our results only strictly apply to long-period orbits, where quadrupole-level expansion of the planetary disturbing function provides an acceptable description of the long-term dynamics. We further remind the reader of the various approximations that we have employed in our formalism. Specifically, we have neglected Neptune's eccentricity along with perturbations arising from the other planets, and have limited our analysis to a common plane. Of course, the solar system is not a 2D restricted three-body problem, meaning that our analytic estimate of the stability boundary is, by construction, inexact. Still, a comparison of our results with direct $N$-body simulations indicates that our estimates are sufficiently close to their numerically computed counterparts to provide a useful blueprint for the dynamical architecture of the distant scattered disk.

We conclude this work by remarking that the stability boundary of the scattered disk does not correspond to a single value of the perihelion distance, as is often quoted in the literature. Instead, for long-period orbits, the critical perihelion distance slowly increases with semi-major axis. In other words, the gravitational ``reach" of Neptune's exterior resonances grows with $a$, such that chaos facilitated by Neptune covers a broader perihelion range at longer periods. Taken in isolation, however, scattered disk objects still obey the conservation of the Tisserand parameter, which is well approximated by the preservation of the perihelion distance for highly eccentric long-period orbits (see Appendix). This means that objects that stochastically diffuse outward through the scattered disk do so at approximately constant $q$, and Neptune scattering alone cannot readily populate the large-$a$ chaotic parameter space with $q\gtrsim36\,$AU. For this reason, the generation of chaotic high-perihelion TNOs must be interpreted as a dynamical signature of the interplay between Neptune's exterior $2:j$ resonances and extrinsic gravitational effects that sculpt the outermost regions of the solar system.


\begin{acknowledgments}
We are indebted to Alessandro Morbidelli, Matt Clement, and Mike Brown for illuminating discussions, as well as to Dan Tamayo for providing a thorough and insightful referee report. We are additionally grateful to Hanno Rein for sharing his expertise in numerical implementation of chaos indicators. K.B. is grateful to Caltech, and the David and Lucile Packard Foundation for their generous support. 
\end{acknowledgments}

\begin{appendix} \label{sec:app}

In the following text, we consider the relationship between the Jacobi constant, the Tisserand parameter, the perihelion distance, and the resonant integral of motion $\Psi$. To start this discussion, let us go back to the full Hamiltonian of the circular planar restricted three-body problem: 
\begin{align}
\Ham=\frac{1}{2}\bigg(P_r^2+\frac{P_{\theta}^2}{r^2} \bigg)-\frac{\G\,\Msun}{r}+V(r,\theta-n_{\rm{p}}\,t)+\mathcal{T},
\end{align}
where $P_r$ is the specific linear momentum conjugate to $r$, $P_{\theta}$ is the specific angular momentum conjugate to the azimuthal angle $\theta$, $\mathcal{T}$ is a dummy action conjugate to $t$, $V$ is the planetary potential, and $n_{\rm{p}}$ is the planetary mean motion. 

\paragraph{The Jacobi Constant} Arguably the most fundamental integral of the restricted three-body problem is the Jacobi constant, which follows directly from the Hamiltonian. Defining a contact transformation through the type-2 generating function $\mathcal{F}_2=(r)\,P_{r}'+(\theta-n_{\rm{p}}\,t)\,P_{\theta}'+(t)\,\Xi$, we have $P_r=P_r'$, $P_{\theta}=P_{\theta}'$, and $\mathcal{T}=\Xi-n_{\rm{p}}\,P_{\theta}$. This canonical change of variables corresponds to a transition into a reference frame that co-rotates with the planet at the orbital frequency $n_{\rm{p}}$, such that the new azimuthal angle is $\theta'=\theta-n_{\rm{p}}\,t$. 

Dropping the new constant dummy action $\Xi$, the Hamiltonian is now expressed as:
\begin{align}
\Ham=\frac{1}{2}\bigg(P_r'^2+\frac{P_{\theta}'^2}{r'^2} \bigg)-\frac{\G\,\Msun}{r'}+V(r',\theta')-n_{\rm{p}}\,P_{\theta}'.
\end{align}
Because $\Ham$ now has no explicit time dependence, it is conserved. This locum of energy in a rotating frame, $\Ham$, \textit{is} the Jacobi constant (technically, the conventional expression of the Jacobi constant differs from $\Ham$ by a factor of $-2$, but this is obviously irrelevant).

\paragraph{The Tisserand Parameter}  The above expression contains one term that is guaranteed to be much smaller than others: by virtue of being proportional to the planet-star mass ratio, $V(r',\theta')$ is assuredly negligible. Accordingly, employing the usual expression for the specific energy of a Keplerian orbit and noting that $P_{\theta}'=\sqrt{\G\,\Msun\,a\,(1-e^2)}$, we obtain:
\begin{align}
\Ham \approx - \frac{\G\,\Msun}{2a}-n_{\rm{p}}\,\sqrt{\G\,\Msun\,a\,(1-e^2)} +\order\bigg(\frac{m_{\rm{p}}}{\Msun} \bigg).
\end{align}
Scaling this expression by the inverse specific energy of the planet, $-2\,a_{\rm{p}}/(\G\,\Msun))$, we obtain the Tisserand parameter:
\begin{align}
T=\alpha+2\sqrt{\frac{1-e^2}{\alpha}}.
\end{align}

\paragraph{The Perihelion Distance} As discussed in the main text of the article, distant scattered disk orbits are characterized by small semi-major axis ratios $\alpha\ll1$ and near-unity eccentricities. Accordingly, expanding the above expression for $T$ to zeroth order in $\alpha$ around $0$ and first order in $e$ around $1$, we obtain
\begin{align}
T\approx2\,\sqrt{2}\,\sqrt{\frac{1-e}{\alpha}}+\order\big(\sqrt{\alpha},(1-e)^{3/2} \big).
\end{align}
Multiplying the square of this approximate expression for the Tisserand parameter by $a_{\rm{p}}/8$, we recover the perihelion distance:
\begin{align}
q=a\,(1-e)\approx\frac{a_{\rm{p}}}{8}\,T^2.
\end{align}

\paragraph{The Resonant Integral} A key consequence of the conservation of the action $\Psi$ (defined in equation \ref{transform}) is that changes in the Delaunay actions $L=\sqrt{\G\,\Msun\,a}$ and $G=\sqrt{\G\,\Msun\,a\,(1-e^2)}$, are related through $\Delta L=\chi\,\Delta G/2$. Let us examine this relationship in further detail. Returning to the ``exact" expression for the Tisserand parameter, let us express it in terms of Delaunay variables: 
\begin{align}
T=\sqrt{\frac{G^2}{\G\,\Msun\,a_{\rm{p}}}}+\frac{\G\,\Msun\,a_{\rm{p}}}{2\,L^2}
\end{align}

Taking the finite difference, we have:
\begin{align}
\sqrt{\frac{1}{\G\,\Msun\,a_{\rm{p}}}}\,\Delta G=\frac{\G\,\Msun\,a_{\rm{p}}}{[L]^3}\,\Delta L.
\end{align}
Rearranging the expression and noting that in the vicinity of a $2:\chi$ resonance $(a/a_{\rm{p}})^{3/2}\approx\chi/2$, we obtain
\begin{align}
\Delta L=\bigg(\frac{a}{a_{\rm{p}}} \bigg)^{3/2}\, \Delta G = \frac{\chi}{2}\,\Delta G.
\end{align}
This is result is identical to the one that ensues from the conservation of $\Psi$, implying that (to within an additive constant) $\Psi$ \textit{is} a near-resonant approximation to the Tisserand parameter.

The above formulae highlight the fact that the approximate maintenance of the perihelion distance by scattered disk objects, the preservation of the resonant action we employed in our analysis, as well as the near-constancy of the Tisserand parameter -- which is itself nothing other than an approximation to the Jacobi constant -- are all re-statements of the same conservation law.

\end{appendix}


\end{document}